\documentclass[a4paper,11pt]{article}
\usepackage{latexsym}
\usepackage{epsfig}
\title{Transition from quintessence to phantom phase in quintom model}
\author{H. Mohseni Sadjadi\footnote{mohsenisad@ut.ac.ir} and M.Alimohammadi\footnote{alimohmd@ut.ac.ir}
\\ {\small School of Physics, University of Tehran,}
\\ {\small North Karegar Ave., Tehran, Iran.}}
\begin{document}
\maketitle

\begin{abstract}
Assuming the Hubble parameter is a continuous and differentiable
function of comoving time, we investigate necessary conditions for
quintessence to phantom phase transition in quintom model. For
power-law and exponential potential examples, we study the
behavior of dynamical dark energy fields and Hubble parameter near
the transition time, and show that the phantom-divide-line
$\omega=-1$ is crossed in these models.
\end{abstract}
\section{Introduction} Astronomical data show that the expansion of
our universe is accelerated at the present epoch \cite{astro}.
Assuming that the universe is filled with perfect fluids, the
equation of state parameter $\omega=p/\rho$ must satisfy
$\omega<-\frac{1}{3}$, indicating a negative pressure. Many
theories have been proposed to study the origin of this negative
pressure or repulsive gravitational behavior.

One of these theories introduces a smooth energy component with
negative pressure dubbed "dark energy". A candidate for dark
energy is the cosmological constant \cite{dark}: a constant
quantum vacuum energy density which fills the space homogeneously,
corresponding to a fluid with a constant equation of state
parameter $\omega=-1$. Because of conceptual problems associated
with the cosmological constant, such as fine-tuning and
coincidence problems \cite{wein}, alternative theories have been
proposed where in a class of them, some dynamical scalar fields,
with suitably chosen potentials, have been introduced to make the
vacuum energy vary with time \cite{dark}. For example to describe
$\omega>-1$ phase or quintessence regime, a normal scalar field
$\phi$, known as quintessence scalar field, can be used
\cite{quintes}. For performing actual calculation, it is useful to
assume that the quintessence field is slowly rolling in a
potential. This approximation (slow-roll), leads to inflationary
expansion of the universe \cite{slow}. $\omega<-1$ phase (or
phantom phase) can be related to the presence of a scalar field
$\sigma$, with a wrong sign kinetic term  dubbed as phantom scalar
field \cite{phan}. Depending on the form of the potential,
different solutions such as asymptotic de Sitter, big rip, etc.
may be obtained \cite{fa}. The effects of gravitational
back-reactions can also counteract that of phantom energy and can
become large enough to end the phantom dominated phase before the
big rip \cite{wu}.

One of the important issues of the scalar models of dark energy is
their stability behaviors in both the classical and quantum
mechanical levels. In classical level, one of the main methods of
stability studies is achieved by studying the stability of the
late-time attractors of the theory by a phase space analysis. This
can be done by determining the eigenvalues of the determinant
obtained by the set of autonomous equations of motion when
perturbed about their critical points. Variety of models have been
studied by this method, which some of them can be found in
\cite{rev} and references therein.

The second kind of instabilities, which is more important in
phantom models, are those based on the quantum fluctuations.
Because the phantom fields have negative kinetic energy, it is
possible that a phantom particle decays into arbitrary number of
phantoms and ordinary particles, such as gravitons. It can be
shown that the decay rates of these interactions are infinite
which indicates that the phantom models are dramatically unstable.
But if we think of these models as the low-energy effective
theories, with the fundamental fields having positive kinetic
energy, then we should use a momentum cutoff $\Lambda$ in
calculating the decay rates. In this way it can be shown that for
$\Lambda \sim M_{\rm pl}$, the lifetimes can become larger than
the age of the universe when one chooses suitable phantom-gravity
interaction potentials, and this removes the quantum instability
of these kinds of phantom models. See \cite{quantum} for a
specific example.

Some present data seems to favor an evolving dark energy with
$\omega$ less than $-1$ at present epoch from $\omega
> -1$ in the near past \cite{Bo1},\cite{Bo2}. It is therefore instructive
to construct physical models in which $\omega$ can cross
$\omega=-1$ line (dubbed as phantom-divide-line). Neither the
cosmological constant nor the dynamical scalar fields, like
quintessence or phantom, can explain the $\omega=-1$ crossing,
hence some other theories have been introduced to describe this
transition \cite{divide}. One of these theories is the quintom
model, which based on hybrid models \cite{linde}\cite{peri1},
describes the transition from $\omega>-1$ to $\omega<-1$ regime by
assuming that the cosmological fluid, besides the ordinary matter
and radiation, is consisted of a quintessence and a phantom scalar
field \cite{Bo1}. A brief introduction of quintom model can be
found in section two. This model can lead to quintessence
domination, i.e. $\omega > -1$, at early time and phantom
domination, i.e. $\omega < -1$, at late time. In \cite{kuo}, a
phase-space analysis of the evolution for a spatially flat
Friedman-Robertson-Walker (FRW) universe containing a barotropic
fluid and phantom-scalar fields with exponential potentials has
been presented. It has been shown that the phantom-dominated
scaling solution is the stable late-time attractor. In
\cite{quintpot} the same calculation has been done by introducing
an interaction term between phantom and quintessence fields.
Recently, the hessence model, as a new view of quintom dark
energy, in which the dark energy is described by a single
non-canonical complex scalar field, rather than two independent
real scalar fields, has been introduced \cite{hessence}. The
evolution of $\omega$ in this model has been studied in \cite{cai}
and \cite{our}, via phase-space analysis. Although this model
allows $\omega$ to cross $-1$, but it avoids the late time
singularity or the "big rip".

In this paper we study the transition from quintessence to phantom
phase in the quintom model. Instead of considering late time
behavior of the model we try to investigate the behavior of
cosmological parameters and scalar fields in a neighborhood of
transition time. In section one, by assuming that the Hubble
parameter is continuous and differentiable function at transition
time, we seek the necessary conditions that must be satisfied in
order that the phase transition occurs. In section two, instead of
a phase-space analysis, we try to obtain solution of Einstein and
Friedmann equations in the vicinity of transition time for
exponential and power law potentials for slowly varying fields. We
show that the dynamical equations are consistent with the
conditions needed for quintessence to phantom phase transition. It
is worth noting that these two specific examples have stable late
time attractor solutions
 \cite{kuo,quintpot}.

Through this paper we use $\hbar=c=G=1$ units.

\section{Phase transition in quintom model: general results}
We consider a spatially flat FRW space-time in comoving
coordinates $(t,x,y,z)$
\begin{equation}\label{1}
ds^2=-dt^2+a^2(t)(dx^2+dy^2+dz^2),
\end{equation}
where $a(t)$ is the scale factor. We assume that the universe is
filled with two kinds of fluid: (dark) matter and quintom dark
energy. The equation of state of matter is
\begin{equation}\label{2}
P_m=(\gamma_m-1)\rho_m, \ \ \ \, 1<\gamma_m<2,
\end{equation}
where $\rho_m$, and $P_m$ are the matter density and pressure,
respectively, and $\gamma_m=1+\omega_m$. The evolution equation of
$\rho_m$ is
\begin{equation}\label{3}
\dot{\rho_m}+3H\gamma_m\rho_m=0.
\end{equation}
$H(t)=\dot{a}(t)/a$ is the Hubble parameter and "dot" denotes time
derivative. The quintom dark energy consists of a negative kinetic
energy scalar field $\sigma$ and a normal scalar field $\phi$,
described by the Lagrangian density \cite{Bo1},\cite{kuo}:
\begin{equation}\label{4}
L_D=\frac{1}{2}\partial_{\mu}\phi\partial^{\mu}\phi-\frac{1}{2}
\partial_{\mu}\sigma\partial^{\mu}\sigma+V(\phi,\sigma),
\end{equation}
where $V(\phi,\sigma)$ is the quintom potential. Restricting
ourselves to homogeneous fields, the energy density $\rho_D$ and
pressure $P_D$ of the homogenous quintom dark energy is then
\begin{eqnarray}\label{5}
\rho_D&=&\frac{1}{2}\dot{\phi}^2-\frac{1}{2}\dot{\sigma}^2+V(\phi,\sigma),\nonumber \\
P_D&=&\frac{1}{2}\dot{\phi}^2-\frac{1}{2}\dot{\sigma}^2-V(\phi,\sigma).
\end{eqnarray}
The evolution equations of the fields are
\begin{eqnarray}\label{6}
&&\ddot{\phi}+3H\dot{\phi}+\frac{\partial V(\phi,\sigma)}{\partial \phi}=0,\nonumber \\
&&\ddot{\sigma}+3H\dot{\sigma}-\frac{\partial
V(\phi,\sigma)}{\partial \sigma}=0.
\end{eqnarray}
Using Einstein equation, one can show that the Hubble parameter
satisfies the Friedmann equations
\begin{eqnarray}\label{7}
H^2&=&\frac{8\pi}{3}(\rho_D+\rho_m)\nonumber \\
&=&\frac{4\pi}{3}[\dot{\phi}^2-\dot{\sigma}^2+2V(\phi,\sigma)+2\rho_m],
\end{eqnarray}
and
\begin{eqnarray}\label{8}
\dot{H}&=&-4\pi(\rho_D+\rho_m+P_D+P_m)\nonumber \\
&=&-4\pi(\dot{\phi}^2-\dot{\sigma}^2+\gamma_m\rho_m).
\end{eqnarray}
Note that eqs.(\ref{6}), (\ref{7}) and, (\ref{8}) are not
independent. The equation of state parameter $\omega$ is defined
through
\begin{equation}\label{9}
\omega=\frac{P_D+P_m}{\rho_D+\rho_m}=\frac{\dot{\phi}^2-\dot{\sigma}^2-2V(\phi,\sigma)
+2(\gamma_m-1)\rho_m
}{\dot{\phi}^2-\dot{\sigma}^2+2V(\phi,\sigma)+2\rho_m }.
\end{equation}
In terms of Hubble parameter we have
\begin{equation}\label{10}
\omega=-1-\frac{2}{3}\frac{\dot{H}}{H^2}.
\end{equation}
For an accelerating universe, $\ddot{a}(t)>0$, we have
$\dot{H(t)}+H^2(t)>0$ and $\omega<-\frac{1}{3}$. Through this
paper we will restrict ourselves to $H(t)>0$. In the quintessence
phase , $\dot{H}<0$ and therefore $\omega>-1$. In the phantom
phase we have $\dot{H}>0$ so $\omega<-1$. In contrast to
cosmological models with only one scalar field, transition from
quintessence to phantom era, in principle, is possible in quintom
model. If $H(t)$ has a local minimum at $t=t_0$, i.e.
$\dot{H}(t_0)=0$, then at $t<t_0$, $\dot{H}(t)<0$ and $\omega
>-1$ and at $t>t_0$, $\dot{H}(t)>0$ and $\omega <-1$. This
behavior is in agreement with present data of equation of state
parameter $\omega$. So the transition from quintessence to phantom
phase, or crossing the phantom-divide-line, can be studied by
investigating the behavior of the Hubble parameter near $t=t_0$
point.

Let $H(t)$ be differentiable on an open set containing $t_0$, then
it can be expanded at $t=t_0$
\begin{equation}\label{11}
H(t)=\sum_{n=0}^\infty\frac{H^{(n)}(t_0)}{n!}(t-t_0)^n \ \ , \ \
\dot{H}(t_0)=0,
\end{equation}
where $H^{(n)}(t_0)$ is the  $n$-th derivative of $H(t)$  at
$t=t_0$. If $\alpha\geq 2$ is the order of the first non-zero
derivative of $H(t)$ at $t=t_0$, then
\begin{equation}\label{12}
H(t)\simeq
h_0+h_1(t-t_0)^{\alpha}+O\left((t-t_0)^{\alpha+1}\right).
\end{equation}
where $h_0=H(t_0)$ and $h_1= \frac{1}{\alpha !}H^{(\alpha)}(t_0)$.
A transition from quintessence to phantom phase occurs when
$\alpha$ is an even integer and $h_1>0$. $h_1>0$ follows from the
fact that $H(t_0)$ must be a local minimum of $H(t)$. If we
considered the transition from phantom to quintessence universe,
we should take $h_1<0$, instead. Using eqs. (\ref {7}), (\ref
{8}), one finds
\begin{equation}\label{13}
H^2=\frac{8\pi}{3}\mathcal{V}-\frac{1}{3}\dot{H},
\end{equation}
where $\mathcal{V}(t)= V(\phi,\sigma)+
(1-\frac{\gamma_m}{2})\rho_m$. By expanding both sides of
eq.(\ref{13}) near $t=t_0$, and noting $H^{(\beta <\alpha
)}(t_0)=0$, one finds
\begin{equation}\label{114}
\mathcal{V}^{(\beta < \alpha -1)}(t_0)=0,
\end{equation}
\begin{equation}\label{14}
h_0^2=\frac{8\pi}{3}\mathcal{V}_0,
\end{equation}
and
\begin{equation}\label{15}
h_1=\frac{8\pi}{\alpha} {\mathcal{V}}_1,
\end{equation}
in which $\mathcal{V}_0=\mathcal{V}(t_0)$ and
$\mathcal{V}_1={1\over (\alpha -1)!}\mathcal{V}^{(\alpha
-1)}(t_0)$. For matter density $\rho_m$, one finds, using
eq.(\ref{3})
\begin{equation}\label{16}
\rho_m^{(\alpha-1)}(t_0)=(-3H(t_0)\gamma_m)^{\alpha-1}\rho_m(t_0),
\end{equation}
so
\begin{equation}\label{17}
h_1=\frac{8\pi}{\alpha!}\left(V^{(\alpha-1)}(\phi(t_0),\sigma(t_0))+
(1-\frac{\gamma_m}{2})(-3\gamma_mh_0)^{\alpha-1}\rho_m(t_0)\right),
\,\alpha\geq 2.
\end{equation}
Assuming $\rho_m(t)>0$ and noting $\gamma_m\leq 2$, the second
term in the right-hand-side of eq.(\ref{17}) is negative for even
$\alpha$'s. So to have $h_1>0$, it is necessary
\begin{equation}\label{18}
V^{(\alpha-1)}(\phi(t_0),\sigma(t_0))>0.
\end{equation}
It is also interesting to note that eq.(\ref{14}) implies
\begin{equation}\label{19}
\mathcal{V}(t_0)> 0.
\end{equation}
But $\mathcal{V}(t)>0$ is a necessary condition for acceleration
of the universe at any time $t$. This can be verified by
considering that when $\omega<-\frac{1}{3}$, eq.(\ref{9}) yields
\begin{equation}\label{20}
\dot{\phi}^2-\dot{\sigma}^2<V(\phi,\sigma)+\left(1-\frac{3}{2}\gamma_m\right)\rho_m.
\end{equation}
Also for $H(t)\neq 0$, eq.(\ref{7}) results
\begin{equation}\label{21}
\dot{\phi}^2-\dot{\sigma}^2>-2V(\phi,\sigma)-2\rho_m.
\end{equation}
$\mathcal{V}(t)>0$ follows from eqs.(\ref{20}) and (\ref{21}).

Let us for a moment relax our even-$\alpha$ condition and look at
the models with odd $\alpha$. Assume that at some time $\tilde{t}$
we have $\dot{H}(t=\tilde{t})=0$. If the system is in quintessence
phase in $t<\tilde{t}$ times, i.e. $\dot{H}(t<\tilde{t})<0$,
expanding $H(t)$ near $t=t_0$ results
\begin{equation}\label{26}
H(t)=h_0+h_1(t-\tilde{t})^{2n+1}+O((t-\tilde{t})^{2n+2}), \, n\geq
1.
\end{equation}
We assume $h_1<0$ to guarantee the existence of quintessence phase
in $t<\tilde{t}$ times (for $h_1>0$, the system describes a
phantom universe). In this case $\tilde{t}$ is an inflection point
and $\dot{H}(t)<0$ for both $t>\tilde{t}$ and $t<\tilde{t}$, and
therefore no transition to phantom phase happens. Assuming the
potential has a lower positive bound $V(\phi,\sigma)\geq v_0>0$,
and using the fact that $\dot{H}(t)<0$ for all $t$'s,
eqs.(\ref{7}) and (\ref{8}) result
\begin{eqnarray}\label{27}
H^2(t)&>&\frac{4\pi}{3}\left((2-\gamma_m)\rho_m+2V(\phi,\sigma)\right)\nonumber
\\
&>&\frac{8\pi}{3}v_0.
\end{eqnarray}
Since $H(t)$ is a decreasing function, the above relation shows
that $H^2(t)$ achieves its minimum at infinity ( provided we
assume that the system remains in quintessence phase for all $t$)
. Thence $\lim_{t\to \infty}{H^2(t)}=(8\pi/3)v_0$, while
$\lim_{t\to \infty}{V(\phi,\sigma)}=v_0$ and $\lim_{t\to
\infty}{\rho_m}=0$, and the system tends to a de Sitter space time
at late time, i.e $\lim_{t\to \infty}{\dot{H}(t)}=0$ \cite{ren}.

Now let us come back to even-$\alpha$ case and study the behavior
of the equation of state parameter $\omega$ near $t=t_0$, where
$\dot{H}(t_0)=0$. For non-zero differentiable $H(t)$, $\omega$ can
be Taylor expanded as (to see a discussion about Taylor expansion
of $\omega$ in terms of scale parameter see \cite{tay})
\begin{equation}\label{28}
\omega(t)=\sum_{n=0}^\infty \frac{\omega^{(n)}(t_0)}{n!}(t-t_0)^n.
\end{equation}
From eq.(\ref{10}), it is found that $\omega(t_0)=-1$ and that the
first non-zero derivative of $\omega(t)$ in its Taylor expansion
is
\begin{equation}\label{29}
\omega^{(\alpha-1)}(t_0)=-\frac{2}{3}\frac{H^{(\alpha)}(t_0)}{H^2(t_0)},
\end{equation}
where as before we have denoted the order of the first non zero
derivative of $H(t)$ at $t=t_0$  by $\alpha$. Therefore
\begin{equation}\label{30}
\omega(t)=-1-\frac{2\alpha}{3}\frac{h_1}{h_0^2}(t-t_0)^{\alpha-1}+O((t-t_0)^{\alpha}),
\end{equation}
or in terms of $\mathcal{V}$,
\begin{equation}\label{31}
\omega(t)=-1-\frac{2\mathcal{V}_1}{\mathcal{V}_0}(t-t_0)^{\alpha-1}+O((t-t_0)^{\alpha}).
\end{equation}
The above equation shows that (i) to cross the phantom-divide-line
$w=-1$ $\alpha$ must be an even integer, (ii) to go from
quintessence phase, $\omega>-1$, to phantom phase, $\omega<-1$, we
must have $h_1>0$, or equivalently $\mathcal{V}_1>0$. Note that if
initially the universe is in phantom phase, the condition (ii) for
phase transition to quintessence era becomes $h_1<0$. These
results are consistent with those we have obtained from studying
the behavior of $H(t)$ near $t=t_0$.

In brief, to have an accelerating universe, a specific combination
of the quintom potential and matter density, denoted by
$\mathcal{V}(t)$, must be positive. Also a transition from
quintessence to phantom phase occurs provided the order of the
first non-vanishing derivative of $H(t)$ at $t=t_0$, in which
$\dot{H}(t_0)=0$, is even and the parameter $h_1$ obtained in
eq.(\ref{17}) is positive.  We will illustrate these general
results and discuss the phase transition via two important
examples: exponential and power like quintom potentials, in the
next section.

\section{Examples}
\subsection{Exponential potential}
As a first example we consider the quintom model with potential
\begin{eqnarray}\label{32}
V&=& V_{\phi}+V_{\sigma} \nonumber \\
&=&v_1e^{\lambda_1\phi}+v_2e^{\lambda_2\sigma}, v_1>0, \,v_2>0.
\end{eqnarray}
To study the possible occurrence of transition from $\omega>-1$ to
$\omega<-1$ for this potential, we try to obtain the solutions of
eqs.(\ref{6}), (\ref{7}), and (\ref{8}), when the Hubble parameter
behaves as
\begin{equation}\label{33}
H(t)=h_0+h_1t^{\alpha}+O(t^{\alpha+1}) ,\, \alpha\geq 2, \,h_1\neq
0,
\end{equation}
near $t=0$, in which $\dot{H(0)}=0$. Eq.(\ref{6}) cannot be solved
exactly with the potential (\ref{32}), but for slowly varying
fields, where $\ddot{\phi}<<H\dot{\phi}$ and
$\ddot{\sigma}<<H\dot{\sigma}$, these equations become
\begin{eqnarray}\label{34}
3(h_0+h_1t^{\alpha})\dot{\phi}(t)&=&-\lambda_1v_1e^{\lambda_1\phi(t)}\\
 \nonumber 3(h_0+h_1t^{\alpha})
\dot{\sigma}(t)&=&\lambda_2v_2e^{\lambda_2\sigma(t)},
\end{eqnarray}
with solutions
\begin{eqnarray}\label{35}
\phi(t)&=&{1\over {\lambda_1}}\ln\left\{{3h_0
\alpha\over{v_1\lambda_1^2\left[ t\Phi({-t^\alpha h_1\over
h_0},1,{1\over \alpha})+c_1h_0 \alpha\right]}}\right\},\\
\nonumber \sigma(t)&=&{1\over {\lambda_2}}\ln\left\{-{3h_0
\alpha\over{v_2\lambda_2^2\left[t\Phi({-t^\alpha h_1\over
h_0},1,{1\over \alpha})+c_2h_0 \alpha\right]}}\right\},
\end{eqnarray}
in which $\Phi (z,a,b)$ is the Lerchphi function. The constants
$c_1$ and $c_2$, in terms of the initial conditions at $t=0$, are
defined through
\begin{eqnarray}\label{36}
\phi(0)&=&{1\over \lambda_1} \ln  \left( {\frac
{3}{v_{{1}}{\lambda_{{1}}}^{2}c_{{1}}}} \right) \\ \nonumber
\sigma(0)&=&{1\over \lambda_2} \ln  \left( -{\frac
{3}{v_{2}{\lambda_{2}}^{2}c_{2}}} \right)
\end{eqnarray}
Hence $c_1>0$ and $c_2<0$. Near $t=0$, up to order O(t), we have
\begin{eqnarray}\label{37}
\ddot{\phi}(0)&=&{1\over {h_0^2c_1^2\lambda_1}} ,\nonumber \\
3h_0\dot{\phi}(0)&=&-{3\over {\lambda_1 c_1}},
\end{eqnarray}
and
\begin{eqnarray}\label{38}
\ddot{\sigma}(0)&=&{1\over {h_0^2c_2^2\lambda_2}}, \nonumber \\
3h_0\dot{\sigma}(0)&=&-{3\over {\lambda_2 c_2}}.
\end{eqnarray}
Therefore the slowly varying condition is equivalent to
\begin{equation}\label{39}
|c_1h_0^2|\gg 1 \ \ \ , \ \ \  |c_2h_0^2|\gg 1.
\end{equation}
These inequalities can be expressed in terms of the potentials as
following:
$|\frac{1}{V_{\phi}}(\frac{dV_{\phi}}{d{\phi}})^2(0)|\ll {h_0^2}$
and $|\frac{1}{V_{\sigma}}(\frac{dV_{\sigma}}{d{\sigma}})^2(0)|\ll
{h_0^2}$, respectively.

$\rho_m(t)$ satisfies the equation
\begin{equation}\label{40}
\dot{\rho}_m(t)+3\gamma_m(h_0+h_1t^{\alpha})\rho_m(t)=0.
\end{equation}
The solution of this equation is
\begin{equation}\label{41}
\rho_m(t)=c_3e^{-3\gamma_m(h_0t+{h_1\over
{\alpha+1}}t^{\alpha+1})},
\end{equation}
where $c_3=\rho_m(t=0)$.

In terms of dimensionless variables $C_1=1/(c_1 h_0^2)$,
$C_2=1/(c_2 h_0^2)$, $C_3=c_3/h_0^2$, $\tau=h_0t$, and
$H_1=h_1/h_0^{\alpha+1}$, eq.(\ref{7}) becomes
\begin{eqnarray}\label{42}
&&
\frac{H^2(\tau)}{h_0^2}=\frac{4\pi}{3}\Bigg\{\frac{\alpha^2}{\left[\tau\Phi(-\tau^\alpha
H_1,1,{1\over \alpha})+{\alpha \over { C_1} }\right]^2
\left(1+H_1\tau^{\alpha}\right)^2\lambda_1^2}-\\
\nonumber &&\frac{\alpha^2}{\left[\tau\Phi(-\tau^\alpha
H_1,1,{1\over \alpha})+{\alpha \over { C_2}
}\right]^2\left(1+H_1\tau^{\alpha}\right)^2\lambda_2^2}+\frac{6\alpha}
{\lambda_1^2\left[\tau\Phi(-\tau^\alpha
H_1,1,{1\over \alpha})+{\alpha \over C_1}\right]}\\
\nonumber &-&\frac{6\alpha}{\lambda_2^2\left[\tau\Phi(-\tau^\alpha
H_1,1,{1\over \alpha})+{\alpha \over
C_2}\right]}+2C_3e^{-3\gamma_m(\tau+\frac{H_1}{\alpha+1}\tau^{\alpha+1})}
\Bigg\},
\end{eqnarray}
and eq.(\ref{8}) reduces to
\begin{eqnarray}\label{43}
&&\frac{1}{h_0}\frac{dH(\tau)}{d\tau}
=-4\pi\Bigg\{\frac{\alpha^2}{\left[\tau\Phi(-\tau^\alpha
H_1,1,{1\over \alpha})+{\alpha \over { C_1}
}\right]^2\left(1+H_1\tau^{\alpha}\right)^2\lambda_1^2}-
\\ \nonumber
&& \frac{\alpha^2}{\left[\tau\Phi(-\tau^\alpha H_1,1,{1\over
\alpha})+{\alpha \over { C_2}
}\right]^2\left(1+H_1\tau^{\alpha}\right)^2\lambda_2^2}+\gamma_m
C_3e^{-3\gamma_m(\tau+\frac{H_1}{\alpha+1}\tau^{\alpha+1})}\Bigg\}.
\end{eqnarray}
By expanding the right hand sides of eqs.(\ref{42}) and (\ref{43})
near $\tau=0$ and using eq.(\ref{33}) for their left hand sides,
one finds:
\begin{eqnarray}\label{45}
&&1+2H_1\tau^{\alpha}+O(\tau^{\alpha+1})=
\frac{4\pi}{3}\left[\frac{C_1(C_1+6)}{\lambda_1^2}-
\frac{C_2(C_2+6)}{\lambda_2^2}+2C_3\right]\nonumber \\
&&-8\pi\left[\frac{C_1^2(\frac{C_1}{3}+1)}{\lambda_1^2}-
\frac{C_2^2(\frac{C_2}{3}+1)}{\lambda_2^2}+\gamma_mC_3 \right]\tau
\nonumber \\
&&+\frac{4\pi}{3}\Bigg[3\left(3\gamma_m^2C_3+2\frac{C_1^3}{\lambda_1^2}-2\frac{C_2^3}{\lambda_2^2}\right)
+2
\left(\frac{C_2^2}{\lambda^2_2}-\frac{C_1^2}{\lambda_1^2}\right)H_1\nonumber
\\ &&+3\left(\frac{C_1^4}{\lambda_1^2}-
\frac{C_2^4}{\lambda_2^2}\right)\Bigg]\tau^2+O(\tau^3),
\end{eqnarray}
and
\begin{eqnarray}\label{44}
&&\alpha H_1\tau^{\alpha-1}+O(\tau^{\alpha})=
-4\pi\left(\gamma_mC_3+\frac{C_1^2}{\lambda_1^2}-\frac{C_2^2}{\lambda_2^2}\right)\nonumber
\\&+&
4\pi\left(3\gamma_m^2C_3+\frac{2C_1^3}{\lambda_1^2}-\frac{2C_2^3}{\lambda_2^2}\right)\tau
+O(\tau^2),
\end{eqnarray}
respectively. To have a consistent set of equations for $\alpha
\geq 2$, one finds the equalities
\begin{equation}\label{46}
\frac{4\pi}{3}\left(\frac{C_1(C_1+6)}{\lambda_1^2}-
\frac{C_2(C_2+6)}{\lambda_2^2}+2C_3\right)=1,
\end{equation}
and
\begin{equation}\label{47}
\frac{C_1^2(\frac{C_1}{3}+1)}{\lambda_1^2}-
\frac{C_2^2(\frac{C_2}{3}+1)}{\lambda_2^2}+\gamma_mC_3 =0,
\end{equation}
from the first two terms of eq.(\ref{45}) and
\begin{equation}\label{48}
\gamma_mC_3+\frac{C_1^2}{\lambda_1^2}-\frac{C_2^2}{\lambda_2^2}=0,
\end{equation}
from the first term of eq.(\ref{44}), respectively. In slow-roll
approximation, which we are working in, the parameters satisfy
eq.(\ref{39}), which imply $|C_1|<<1$ and $|C_2|<<1$. In this
approximation, the eqs.(\ref{47}) and (\ref{48}) are equivalent.

The important observation is that since $C_1$ and $C_3$ are
positive and $C_2$ is negative real numbers, the coefficient of
$\tau$ in eq.(\ref{44}) is a non-zero positive number, which
results the following for exponential potential:
\begin{eqnarray}\label{146}
\alpha_{\rm exp.}&=&2,\nonumber \\ (h_1)_{\rm exp.}&>&0.
\end{eqnarray}
In this way it is proved that in the quintom model with
exponential potential, the system has a phase transition from
quintessence to phantom phase, if any other remaining relations,
obtained from eqs.(\ref{45}) and (\ref{44}), will be satisfied
consistently up to the lowest order.

To check the remaining relations, we first note that since $\alpha
=2$, eq.(\ref{44}) results another equation, as follows,
\begin{equation}\label{49}
H_1=2\pi\left(3\gamma_m^2C_3+\frac{2C_1^3}{\lambda_1^2}-\frac{2C_2^3}{\lambda_2^2}\right).
\end{equation}
But eqs.(\ref{47}) and (\ref{48}) imply: $C_1^3/\lambda_1^2-
C_2^3/\lambda_2^2<<|C_1^2/\lambda_1^2-C_2^2/\lambda_2^2|\simeq
C_3$, so eq.(\ref{49}) reduces to:
\begin{equation}\label{148}
H_1\simeq 6\pi \gamma_m^2C_3.
\end{equation}
This shows that the rate of the phase transition depends on the
parameter of state $\gamma_m$ and the dark matter density at
transition time.

In terms of dark energy and matter densities, $\rho_D$, $\rho_m$,
eq.(\ref{46}) can be written as
\begin{equation}\label{50}
\frac{\rho_m}{\rho_c}+\frac{\rho_D}{\rho_c}=1,
\end{equation}
where $\rho_c=3h_0^2/8\pi$, $\rho_m(0)/\rho_c=(8\pi/3)C_3$ and
$\rho_D(0)/\rho_c=(4\pi/3)[C_1(C_1+6)/\lambda_1^2-C_2(C_2+6)/\lambda_2^2]$.
The dark energy density is composed of two parts: the density of
kinetic dark energy,
$(4\pi/3)(C_1^2/\lambda_1^2-C_2^2/\lambda_2^2)\rho_c$, which is
negative according to eq.(\ref{48}), and the density of potential
dark energy, $(4\pi/3)(6C_1/\lambda_1^2-6C_2/\lambda_2^2)\rho_c$.
Now since $C_1/\lambda_1^2-
C_2/\lambda_2^2>>C_1^2/\lambda_1^2-C_2^2/\lambda_2^2\simeq C_3$,
it is obvious that the main part of the energy density at the
transition time is coming from the potential of dark energy.

The remaining equation that can be extracted from eq.(\ref{45}) is
\begin{eqnarray}\label{51}
H_1&=&\frac{2\pi}{3}\Bigg[3\left( 3\gamma_m^2C_3+2\frac{C_1^3}
{\lambda_1^2}-2\frac{C_2^3}{\lambda_2^2}\right) +2
\left(\frac{C_2^2}{\lambda^2_2}-\frac{C_1^2}{\lambda_1^2}\right)H_1\nonumber
\\
&&+3\left(\frac{C_1^4}{\lambda_1^2}-
\frac{C_2^4}{\lambda_2^2}\right)\Bigg]\nonumber \\ &=& H_1\Bigg[
1+ 2 \left(\frac{C_2^2}{\lambda^2_2}-\frac{C_1^2}{\lambda_1^2}
\right)+O(C^3)\Bigg],
\end{eqnarray}
which can be verified using the fact that
$C_1^2/\lambda_1^2-C_2^2/\lambda_2^2 <<
C_1/\lambda_1^2-C_2/\lambda_2^2\simeq \rho_{\rm total}/\rho_c=1$.
This completes our desired checking of all relations in the lowest
order.

Expanding
$\mathcal{V}(t)/h_0^2=(V+(1-\frac{\gamma_m}{2})\rho_m)/h_0^2$ at
$t=0$ results
\begin{eqnarray}\label{52}
\frac{\mathcal{V}(t)}{h_0^2}&=&\frac{(2-\gamma_m)C_3}{2}
+\frac{3C_1}
{\lambda^2 _1}- \frac{3C_2}{\lambda_2^2}\nonumber \\
&&+\frac{3}{2}\left[C_3\left(\gamma_m^2-2\gamma_m\right)+
\frac{2C_2^2}{\lambda_2^2}-
\frac{2C_1^2}{\lambda_1^2}\right]\tau+O(\tau^2)\nonumber \\
&=&\frac{3}{8\pi}+\frac{1}{4\pi}H_1\tau + O(\tau^2),
\end{eqnarray}
where we have used eqs.(\ref{46}), (\ref{48}), and (\ref{148}) in
the second equality. The above result is in complete agreement
with our previous assertions eqs.(\ref{14}) and (\ref{15}).

\begin{figure}
\centering\epsfig{file=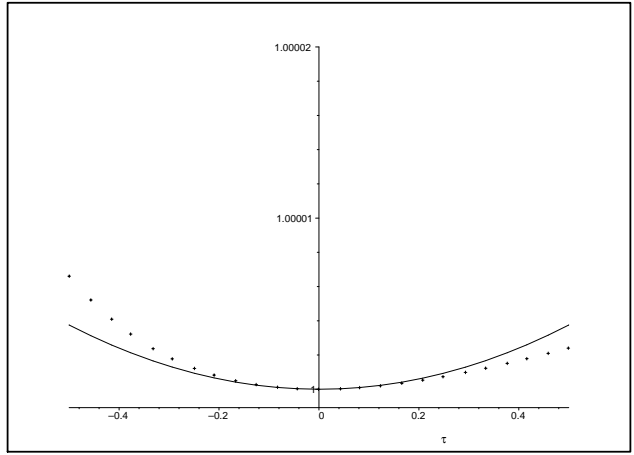,width=6cm,angle=270}
\caption{$H^2/h_0^2$ as a function of $\tau$, obtained from
eq.(\ref{42}) (points), and $(1+H_1\tau^2)^2$ (line).}
\label{fig1}
\end{figure}

\begin{figure}\centering\epsfig{file=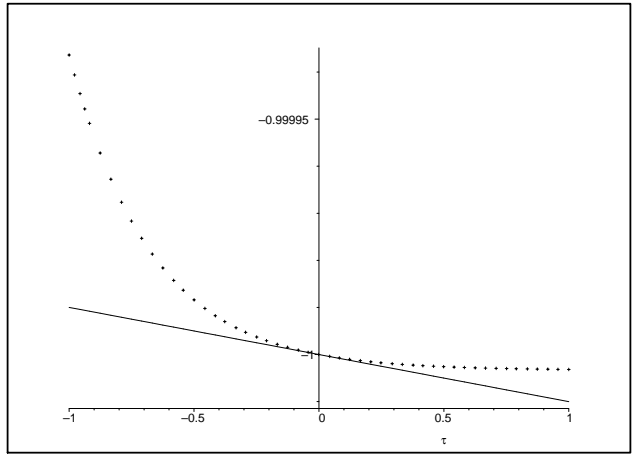,width=6cm,angle=270} \caption{
$\omega$ versus $\tau$, using approximation (\ref{42}) (points)
and $H=h_0(1+H_1\tau^2)$ (line) in eq.(\ref{10}).} \label{fig2}
\end{figure}

To test, in another way, the validity of our approximations let us
choose $C_1=10^{-4}$, $C_2=-1.1\times 10^{-4}$ ( since it must be
small quantities ), $\gamma_m=1$, and $\lambda_1=\lambda_2$ ( for
simplicity ), at $t=0$. Eqs.(\ref{46}), (\ref{48}) and (\ref{49})
then result $C_3=4.0\times 10^{-7}$, $\lambda_1=\lambda_2=0.073$,
and $H_1=7.5\times 10^{-6}$. Fig.(\ref{fig1}) shows the plot of
$(1+H_1\tau^2)^2$ and the same plot obtained from eq.(\ref{42}).
Fig.(\ref{fig2}) shows the comparison of $\omega$ in these two
approaches.

At the end, it is worth noting that all the above approximations
are valid until $\tau <<1$, which is equivalent to $t<<h_0^{-1}$.
But this is reasonable period of time since $h_0^{-1}$ is of order
of the age of our universe.

\subsection{Power-law potential}
As the second example consider the potential
\begin{eqnarray}\label{53}
V&=&V_{\phi}+V_{\sigma}\nonumber \\
 &=&v_1\phi^{b_1}+v_2\sigma^{b_2},
\end{eqnarray}
Using the approximation (\ref{33}), the solutions of the
eq.(\ref{6}) are
\begin{eqnarray}\label{54}
\phi(t)&=&\left(\frac{v_1b_1(b_1-2)t\Phi(\frac{-t^{\alpha}h_1}{h_0},1,\frac{1}{\alpha})}
{3h_0\alpha}+c_1  \right)^{\frac{1}{2-b_1}}\nonumber \\
\sigma(t)&=&\left(\frac{v_2b_2(-b_2+2)t\Phi(\frac{-t^{\alpha}h_1}{h_0},1,\frac{1}{\alpha})}
{3h_0\alpha}+c_2  \right)^{\frac{1}{2-b_1}}
\end{eqnarray}
where $c_1=\phi^{2-b_1}(0)$ and $c_2=\sigma^{2-b_2}(0)$. We assume
$\phi(0)>0$ and $\sigma(0)>0$. Using these solutions one can show
that neglecting second derivatives in eq.(\ref{6}), is allowed
when
\begin{eqnarray}\label{55}
&&|b_1(b_1-1)\frac{v_1}{h_0^2}|\phi^{b_1-2}(0)\ll 1, \nonumber \\
&&|b_2(b_2-1)\frac{v_2}{h_0^2}|\sigma^{b_2-2}(0)\ll 1.
\end{eqnarray}
In term of potentials, these inequalities are
$|\frac{d^2V_\phi}{d\phi^2}|(0)\ll h_0^2$ and
$|\frac{d^2V_\sigma}{d\sigma^2}|(0)\ll h_0^2$. For $b_1=b_2=2$;
$v_1=m_\phi^2/2$ and $v_2=m_\sigma^2/2$, eq.(\ref{54}) reduces to
\begin{eqnarray}\label{56}
\phi(t)&=&\phi(0) e^{-\frac{m_\phi^2t\Phi \left (-\frac{t^\alpha
h_1}{h_0},1,\frac{1}{\alpha}\right)}
{3h_0\alpha}}  \nonumber \\
\sigma(t)&=&\sigma(0)e^{\frac{m_\sigma^2t\Phi
\left(-\frac{t^{\alpha}h_1}{h_0},1,\frac{1}
{\alpha}\right)}{3h_0\alpha}}.
\end{eqnarray}
In this case, the approximation (\ref{55}) is reduced to small
mass limits for phantom and quintessence fields: $m_\phi\ll h_0$
and $m_\sigma\ll h_0$.

$\rho_m$ is given by eq.(\ref{41}). By putting eqs.(\ref{41}) and
(\ref{54}) into eqs.(\ref{7}) and (\ref{8}), and using
dimensionless variables $V_1=v_1/h_0^2$, $V_2=v_2/h_0^2$,
$C_3=c_3/h_0^2$, $\tau=th_0$, and $H_1=h_1/h_0^{\alpha+1}$, near
$\tau=0$,  we obtain
\begin{eqnarray}\label{58}
&&1+2H_1\tau^{\alpha}+O(\tau^{\alpha+1})=
\frac{4\pi}{3}\Bigg[\frac{\phi^{2b_1-2}(0)b_1^2V_1^2}{9}-
\frac{\sigma^{2b_2-2}(0)b_2^2V_2^2}{9}+
2C_3+2V_1\phi^{b_1}(0)+\nonumber
\\
&&
2V_2\sigma^{b_2}(0)\Bigg]+\frac{4\pi}{3}\Bigg[-\frac{2b_1^3(b_1-1)V_1^3\phi^{3b_1-4}(0)}{27}-
\frac{2b_2^3(b_2-1)V_2^3\sigma^{3b_2-4}(0)}
{27}-6\gamma_mC_3\nonumber
\\
&&-\frac{2}{3}(V_1^2b_1^2\phi^{2(b_1-1)}(0)-V_2^2b_2^2\sigma^{2(b_2-1)}(0))\Bigg]\tau\nonumber \\
&& +\frac{4\pi}{3}\Bigg[3\left(\frac{2b_1^3(b_1-1)\phi^{3b_1-4
}(0)V_1^3}{27}+\frac{2b_2^3(b_2-1)\sigma^{3b_2-4
}(0)V_2^3}{27}+3\gamma^2_mC_3\right)\nonumber \\
&&+2\left(-\frac{\phi^{2(b_1-1)}(0)b_1^2V_1^2}{9}+
\frac{\sigma^{2(b_2-1)}(0)b_2^2V_2^2}{9}\right)H_1+
\frac{b_1^4(3b_1^2-7b_1+4)V_1^4\phi^{4b_1-6}(0)}{81}\nonumber \\
&&+
\frac{b_2^4(3b_2^2-7b_2+4)V_2^4\sigma^{4b_2-6}(0)}{81}\Bigg]\tau^2+O(\tau^3),
\end{eqnarray}
and
\begin{eqnarray}\label{57}
&&\alpha H_1\tau^{\alpha-1}+O(\tau^{\alpha})
=-4\pi\left[\frac{\phi^{2(b_1-1)}(0)b_1^2V_1^2}{9}-
\frac{\sigma^{2(b_2-1)}(0)b_2^2V_2^2}{9}+
\gamma_mC_3\right]\nonumber \\
&&-4\pi\left[-\frac{2b_1^3(b_1-1)\phi^{3b_1-4}(0)V_1^3}{27}-
\frac{2b_2^3(b_2-1)\sigma^{3b_2-4}(0)V_2^3}{27}-3\gamma_m^2C_3\right]
\tau\nonumber \\ &&+O(\tau^2),
\end{eqnarray}
respectively. Therefore if the assumption (\ref{33}) is true, we
must have :
\begin{equation}\label{59}
\frac{\phi^{2(b_1-1)}(0)b_1^2V_1^2}{9}-
\frac{\sigma^{2(b_2-1)}(0)b_2^2V_2^2}{9}+ \gamma_mC_3=0,
\end{equation}
\begin{eqnarray}\label{60}
&&\gamma_mC_3+\frac{1}{9}\Bigg[V_1^2b_1^2\phi^{2(b_1-1)}(0)\left(1+\frac{1}{9}
b_1(b_1-1)V_1\phi^{b_1-2}(0)\right)-\nonumber \\
&&V_2^2b_2^2\sigma^{2(b_2-1)}(0)\left(1-\frac{1}{9}b_2(b_2-1)V_2\phi^{b_2-2}(0)\right)\Bigg]=0,
\end{eqnarray}
and
\begin{equation}\label{61}
\frac{4\pi}{3}\left[\frac{\phi^{2b_1-2}(0)b_1^2V_1^2}{9}-
\frac{\sigma^{2b_2-2}(0)b_2^2V_2^2}{9}+2C_3+2V_1\phi^{b_1}(0)+2V_2\sigma^{b_2}(0)\right]=1.
\end{equation}
For slowly varying fields (eq.(\ref{55})), eq.(\ref{60}) reduces
to eq.(\ref{59}). The coefficient of $\tau$ in eq.(\ref{57}) is
\begin{eqnarray}\label{62}
&&4\pi \Big[\frac{2b_1^3(b_1-1)\phi^{3b_1-4}(0)V_1^3}{27}+
\frac{2b_2^3(b_2-1)\sigma^{3b_2-4}(0)V_2^3}{27} \nonumber \\
&&+3\gamma_m^2C_3\Big].
\end{eqnarray}
Comparing eqs.(\ref{59}) and (\ref{60}) shows
$\phi^{3b_1-4}(0)V_1^3+\sigma^{3b_2-4}(0)V_2^3 <<
\phi^{2b_1-2}(0)V_1^2-\sigma^{2b_2-2}(0)V_2^2 \simeq C_3$, which
reduces eq.(\ref{62}) to
\begin{equation}\label{63}
H_1\simeq 6\pi\gamma_m^2C_3.
\end{equation}
This is a non-zero positive number. So for quintom model with
power-law potential, we show
\begin{eqnarray}\label{163}
\alpha_{\rm power-law}&=&2,\nonumber \\ (h_1)_{\rm power-law}&>&0,
\end{eqnarray}
which proves the existence of quintessence to phantom phase
transition in this model. Of course the remaining relations must
be also checked. Note that the value of $H_1$ at transition point
does not depend on the potential, compare eqs.(\ref{148}) and
(\ref{63}), a fact that we guess is true for any other potential.

To check the consistency of the remaining equations, first we note
that by the same arguments used in the preceding example, it can
be shown that the main contribution in eq.(\ref{61}), which can be
written as $\rho_{\rm total}/\rho_c=1$, comes from the quintom
potential, i.e. $V_1\phi^{b_1}(0)+V_2\sigma^{b_2}(0)$. Now the
coefficient of $\tau^2$ in the right-hand-side of eq.(\ref{58}),
using eq.(\ref{62}), can be written as
\begin{equation}\label{64}
2H_1\left[1+\frac{4\pi}{3}\left(-\frac{\phi^{2(b_1-1)}(0)b_1^2V_1^2}{9}+
\frac{\sigma^{2(b_2-1)}(0)b_2^2V_2^2}{9}\right)+\cdots \right],
\end{equation}
which is equal to $2H_1$, using the fact that
$-V_1^2\phi^{2(b_1-1)}(0)+V_2^2\sigma^{2(b_2-1)}(0)<<
V_1\phi^{b_1}(0)+V_2\sigma^{b_2}(0)\simeq \rho_{\rm
total}/\rho_c=1$. Finally it can be shown that eq.(\ref{52}) is
also verified for this potential.

As an illustration of the transition behavior of quintom model
with potential (\ref{53}), see Fig(3), which shows the plot of
$H^2/h_0^2$ obtained from eq.(\ref{7}), for $b_1=b_2=2,$ $
\gamma_m =1$, $ v_1=m_\phi^2/2,\, v_2=m_\sigma^2/2$, with
$m_\phi/h_0=m_\sigma /h_0=0.01$ ( since they must be small
quantities ) and $ \sigma^2(0)=2\phi^2(0)$. Eqs. (\ref{59}),
(\ref{61}) and (\ref{63}) then result $\phi^2(0) = 2500.0/\pi$, $
C_3 =0.28\times 10^{-5}/\pi$, and $ H_1 = 0.17\times 10^{-4}$. See
also Fig.(\ref{fig4}) for $\omega$.
\begin{figure}
\centering\epsfig{file=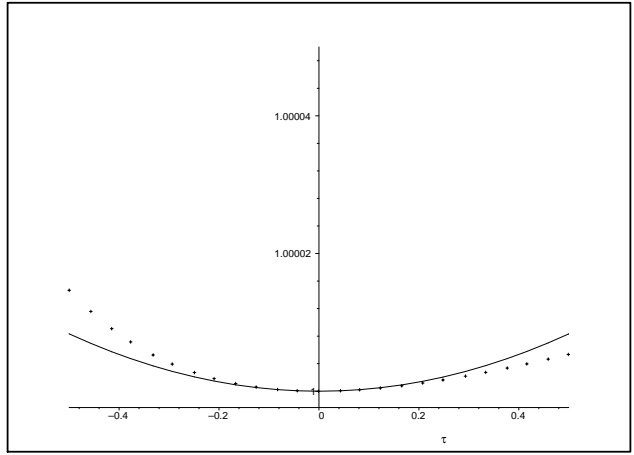,width=6cm,angle=270}
\caption{$H^2/h_0^2$ as a function of $\tau$, obtained from
eq.(\ref{7}) (with initial conditions mentioned in the text)
(points), and $(1+H_1\tau^2)^2$ (line).} \label{fig3}
\end{figure}
\begin{figure}\centering\epsfig{file=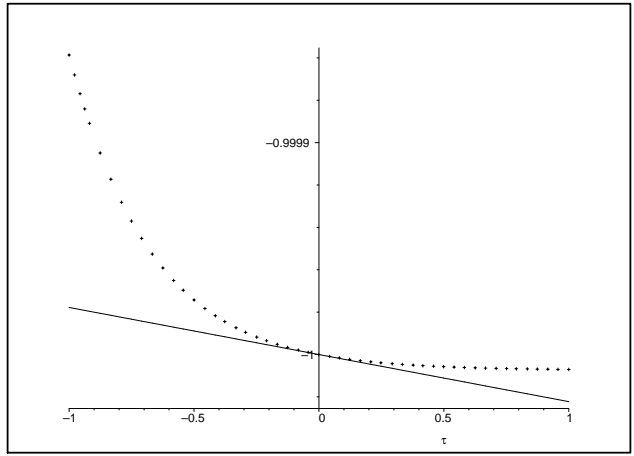,width=6cm,angle=270} \caption{
$\omega$ versus $\tau$, using approximation (\ref{58}) (points)
and $H=h_0(1+H_1\tau^2)$ (line) in eq.(\ref{10}) for power-law
potential.} \label{fig4}
\end{figure}
\section{Conclusion}
In this paper the phase transition from quintessence to phantom
era in quintom model has been discussed. By the assumption that
the Hubble parameter has a Taylor expansion in terms of comoving
time, the behavior of the Hubble parameter and the potential near
the transition time, and the relation between them, have been
studied. See eqs.(\ref{114})-(\ref{15}). The conditions need to
satisfy to have a quintessence to phantom phase transition have
been calculated, i.e. evenness of $\alpha$ and eq.(\ref{17}). The
same results obtained by studying the relations between equation
of state parameter $\omega$, the quintom potential and matter
density, see eq.(\ref{31}).

To be specific, we have considered special cases by determining
the quintom potential. For slowly varying exponential quintom
potential, we have shown that the equations are consistent when
$\alpha=2$ and $h_1>0$. In this way it has been proved that the
phase transition occurs with a rate depending on the density of
matter at transition time, see the discussion after
eq.(\ref{148}). The relation of matter and dark energy fields and
their densities have been obtained at transition time. See
eqs.(\ref{46}), (\ref{48}) and discussion after eq.(\ref{50}). For
slowly varying power law potentials, similar results has been
deduced.

 In these examples we restricted ourselves to slow roll
quintom model near the transition time. For $\gamma_m\simeq 1$, we
expect at the transition time $\dot{H}=0$, the kinetic energy of
quintom  field be of the same order of magnitude as dark matter
density, see eq.(\ref{8}). Using this conclusion in eq.(\ref{7})
results that the main part of the energy density is provided by
the potential energy of scalar fields. Therefore the dark matter
density is very small with respect to the density of dark energy,
hence the coincidence problem still unsolved in these examples.

{\bf Acknowledgement:} We would like to thank the center of
excellence of the School of Physics of the University of Tehran
for partial financial support.


\begin{thebibliography}{99}
\bibitem{astro} S. Perlmutter et al, Nature (London) {\bf 391}, 51(1998);
D. N. Spergel et al. [WMAP Collaboration], Astrophys. J. Suppl.
{\bf 148}, 175 (2003); A. C. S. Readhead et al., Astrophys. J.
{\bf 609}, 498 (2004);  J. H. Goldstein et al., Astrophys. J. {\bf
599}, 773 (2003);  M. Tegmark et al. [SDSS Collaboration], Phys.
Rev. D {\bf 69}, 103501 (2004);  A. G. Riess et al. [Supernova
Search Team Collaboration], Astron. J. {\bf 116}, 1009 (1998); S.
Perlmutter et al. [Supernova Cosmology Project Collaboration],
Astrophys. J. {\bf 517}, 565 (1999).
\bibitem{dark} T. Padmanabhan, Phys. Rept. {\bf 380}, 235 (2003) ;
V. Sahni and A. A. Starobinsky, Int. J. Mod. Phys. D {\bf 9}, 373
(2000) ; S. M. Carroll, Living Rev. Rel. {\bf 4}, 1 (2001).
\bibitem{wein} S. Weinberg, Rev. Mod. Phys. {\bf 61}, 1 (1989);
I. Zlatev, L.-M. Wang, and P. J. Steinhardt, Phys. Rev. Lett. {\bf
82}, 896 (1999).
\bibitem{quintes}R. R. Caldwell, R. Dave and P. J. Steinhardt,
Phys. Rev. Lett. {\bf 80}, 1582 (1998); P. J. E. Peebles and A.
Vilenkin, Phys. Rev. D {\bf 59}, 063505 (1999) ; P. J. Steinhardt,
L. M. Wang and I. Zlatev, Phys. Rev. D {\bf 59}, 123504 (1999); M.
Doran and J. Jaeckel, Phys. Rev. D {\bf 66}, 043519 (2002).
\bibitem{slow}A. R. Liddle, P.  Parson and J. D. Barrow, Phys. Rev.
D {\bf 50}, 7222 (1994).
\bibitem{phan}R. R. Caldwell, Phys. Lett. B {\bf 545}, 23 (2002);
R. R. Caldwell, M. Kamionkowski and N. N.Weinberg, Phys. Rev.
Lett. {\bf 91}, 071301 (2003); S. M. Carroll, M. Hoffman and M.
Trodden, Phys. Rev. D {\bf 68}, 023509 (2003); J. M. Cline, S. Y.
Jeon and G. D. Moore, Phys. Rev. D {\bf 70}, 043543 (2004).
\bibitem{fa}V. Faraoni, Class. Quant. Grav. {\bf 22}, 3235 (2005).
\bibitem{wu}P. Wu and H. Yu, Nucl. Phys. B {\bf 727}, 355 (2005); gr-qc/0604117.
\bibitem{rev}E. J. Copeland, M. Sami and S. Tsujikawa,
 hep-th/0603057.
\bibitem{quantum}S. M. Carroll, M. Hoffman and M. Trodden, Phys.
Rev. D {\bf 68}, 023509 (2003).
\bibitem{Bo1}B. Feng, X. L. Wang and X. M. Zhang, Phys. Lett. B {\bf 607},
35 (2005).
\bibitem{Bo2} D. Huterer and A. Cooray, Phys. Rev. D
{\bf 71}, 023506 (2005); S. Nesseris and L. Perivolaropoulos,
Phys. Rev. D {\bf 72}, 123519 (2005); U. Seljak, A. Slosar and P.
McDonald, astro-ph/0604335.
\bibitem{divide}Y. H. Wei and Y. Z. Zhang, Grav. Cosmol. {\bf 9}, 307 (2003);
Y. H. Wei and Y. Tian, Class. Quant. Grav. {\bf 21}, 5347 (2004);
R. G. Cai, H. S. Zhang and A. Wang, Commun. Theor. Phys. {\bf 44},
948 (2005); V. Sahni and Y. Shtanov, JCAP {\bf 0311}, 014 (2003);
I. Y. Arefeva, A. S. Koshelev and S. Y. Vernov, Phys. Rev. D {\bf
72}, 064017 (2005); A. Vikman, Phys. Rev. D {\bf 71}, 023515
(2005); A. Anisimov, E. Babichev and A. Vikman, JCAP {\bf 0506},
006 (2005); B. Wang, Y. G. Gong and E. Abdalla, Phys. Lett. B {\bf
624}, 141 (2005); F. C. Carvalho and A. Saa, Phys. Rev. D {\bf
70}, 087302 (2004); H. Mohseni Sadjadi, Phys. Rev. D {\bf 73},
063525 (2006); W. Zhao and Y. Zhang, Phys. Rev. D {\bf 73}, 123509
(2006); P. S. Apostolopoulos and  N. Tetradis, hep-th/0604014; I.
Ya. Aref'eva and A. S. Koshelev, hep-th/0605085; S. Nojiri and
S.D. Odintsov, hep-th/0506212; S. Nojiri, S. D. Odintsov and S.
Tsujikawa, hep-th/0501025; F. Piazza and S. Tsujikawa, JCAP {\bf
0407}, 004 (2004).

\bibitem{linde}A. Linde, Phys. Rev. D {\bf 49}, 748 (1994).
\bibitem{peri1} L. Perivolaropoulos, Phys. Rev. D {\bf 71}, 063503
(2005).
\bibitem{kuo}Z. Guo, Y. Piao, X. M. Zhang and Y. Z. Zhang, Phys. Lett. B {\bf
608}, 177 (2005).
\bibitem{quintpot}X. F. Zhang, H. Li, Y. Piao and X. M. Zhang, Mod. Phys. Lett. A {\bf
21}, 231 (2006); R. Lazkoz and G. Leon,  Phys. Lett. B {\bf 638},
303 (2006).
\bibitem{hessence}H. Wei, R. G. Cai and D. F. Zeng, Class. Quant. Grav. {\bf 22}, 3189
(2005).
\bibitem{cai}H. Wei and R. G. Cai, Phys.
Rev. D {\bf 72}, 123507 (2005).
\bibitem{our}M. Alimohammadi and H. Mohseni Sadjadi, Phys. Rev. D {\bf 73}, 083527
 (2006).
\bibitem{ren}A. D. Rendall, Class. Quant. Grav. {\bf 21}, 2445 (2004); ibid.
 {\bf 22}, 1655 (2005).
\bibitem{tay} W. F. Kao, gr-qc/0605047.




\end{thebibliography}
\end{document}